\documentclass[aps, prl, reprint, nofootinbib,superscriptaddress]{revtex4-1}
\pdfoutput=1
\usepackage{amsmath}
\usepackage{amsfonts}
\usepackage{amssymb}
\usepackage{graphicx}
\usepackage{lmodern}

\newcommand{\doctit}{Charm decay in slow-jet supernovae as the origin of the IceCube ultra-high energy neutrino events}
\newcommand{\doctitle}{Charm decay in slow-jet supernovae as the origin of the IceCube\\ ultra-high energy neutrino events}
\newcommand{\auth}{Atri Bhattacharya, Rikard Enberg, Mary Hall Reno, Ina Sarcevic}

\usepackage[
            pdftitle={\doctit},
            pdfkeywords={IceCube, Supernova, Ultra-high energy neutrinos},
            pdfauthor={\auth},
           ]{hyperref}
\usepackage[all]{hypcap}
\usepackage{array}
\usepackage{slashed} 
\usepackage{hyphenat}
\usepackage[normalem]{ulem}
\usepackage{dcolumn} 
\usepackage{paralist}
\usepackage{mathrsfs} 

\newcommand{\bigabs}[1]{\bigg\lvert #1 \bigg\rvert}

\newcommand{\eg}{\textit{e.g.}}

\newcommand{\anti}[1]{\ensuremath{\bar{#1}}}

\newcommand{\OM}{\ensuremath{\Omega_\text{M}}}
\newcommand{\OLam}{\ensuremath{\Omega_{\Lambda}}}

\newcommand{\order}[1]{\ensuremath{\mathcal{O}(#1)}}

\newcommand{\nue}{\ensuremath{\nu_e}}
\newcommand{\numu}{\ensuremath{\nu_\mu}}
\newcommand{\nutau}{\ensuremath{\nu_\tau}}
\newcommand{\antinue}{\ensuremath{\bar{\nu}_e}}
\newcommand{\antinumu}{\ensuremath{\bar{\nu}_\mu}}
\newcommand{\antinutau}{\ensuremath{\bar{\nu}_\tau}}

\newcommand{\ic}{\ensuremath{\text{IceCube}}}
\newcommand{\cmssr}{\ensuremath{\text{ cm}^{-2}\text{ s}^{-1}\text{ sr}^{-1}}}
\newcommand{\flxunit}{\ensuremath{\text{ GeV}\cmssr}}

\newcommand{\ud}{\ensuremath{\mathrm{d}}}

\newcommand{\diff}{\ensuremath{\,\ud}}

\newcommand{\dnot}{\ensuremath{D^{0}}}
\newcommand{\dnotb}{\ensuremath{\bar{D}^{0}}}
\newcommand{\dpm}{\ensuremath{D^{\pm}}}

\newcommand{\hubbleunit}{\ensuremath{\text{ km s}^{-1}\text{ Mpc}^{-1}}}
\newcommand{\gamj}{\ensuremath{\Gamma_{j}}}
\newcommand{\ndotsn}{\ensuremath{\dot{n}_\text{sn}}}
\newcommand{\sfr}{SFR}
\newcommand{\xisn}{\ensuremath{\xi_\text{sn}}}
\newcommand{\dmess}{$D$-mesons}

\begin{document}

\title{\doctitle}

\author{Atri Bhattacharya}
\email{atrib@email.arizona.edu}
\affiliation{Department of Physics, University of Arizona, Tucson, AZ 85721, USA}
\author{Rikard Enberg}
\email{rikard.enberg@physics.uu.se}
\affiliation{Department of Physics and Astronomy, Uppsala University, Box 516, 751 20 Uppsala, Sweden}
\author{Mary Hall Reno}
\email{mary-hall-reno@uiowa.edu}
\affiliation{Department of Physics and Astronomy, University of Iowa, Iowa City, IA 52242, USA}
\author{Ina Sarcevic}
\email{ina@physics.arizona.edu}
\affiliation{Department of Physics, University of Arizona, Tucson, AZ 85721, USA}
\affiliation{Department of Astronomy and Steward Observatory, University of Arizona, Tucson, AZ 85721, USA}

\begin{abstract}
We investigate whether the recent ultra-high energy (UHE) neutrino events detected at the IceCube neutrino observatory could come from the decay of charmed mesons produced within the mildly relativistic jets of supernova-like astrophysical sources.
We demonstrate that the $5.7\sigma$ excess of neutrinos observed by IceCube in the energy range 30 TeV--2 PeV can be explained by a diffuse flux of neutrinos produced in such slow-jet supernovae, using the values of astrophysical and QCD parameters within the theoretical uncertainties associated with neutrino production from charmed meson decay in astrophysical sources.
We discuss the theoretical uncertainties inherent in the evaluation of charm production in high energy hadronic collisions, as well as the astrophysical uncertainties associated with slow-jet supernova sources.
The proton flux within the source, and therefore also the produced neutrino flux, is cut off at around a few PeV, when proton cooling processes become dominant over proton acceleration. This directly explains the sudden drop in event rates at energies above a few PeV.
We incorporate the effect of energy dependence in the spectrum-weighted charm production cross-section and show that this has a very significant effect on the shape, magnitude and cut-off energies for the neutrino flux.
\end{abstract}

\maketitle

\section{\label{sec:intro}Introduction}

Recently the  IceCube neutrino telescope at the South Pole  has reported the observation of 37 neutrino events in the energy range 30 TeV--2.1 PeV, accumulated over three years of runtime \cite{Aartsen:2013bka,Aartsen:2013jdh,Aartsen:2014gkd}.
These events are $5.7\sigma$ above the atmospheric neutrino background, and present, possibly, the first observation of astrophysical neutrinos.
The reconstructed flux-spectrum from these events suggest conformity with an isotropic $E^{-2}$ spectrum up to  energies of $\sim 2$ PeV, with the best-fit per-flavor $\nu + \anti{\nu}$ flux in this energy range being given by
\begin{equation}\label{eqn:ic-bf}
E^{2}\Phi = (0.95 \pm 0.3) \times 10^{-8} \flxunit.
\end{equation}
The ANTARES neutrino telescope, meanwhile, reports an upper limit on this flux of $E^{2}\Phi = 4.8 \times 10^{-8} \flxunit$ at 90\% confidence level \cite{VanElewyck:2013xja}.
The \ic\ measurement is, in principle, consistent with the theoretical expectations for a diffuse neutrino flux from extragalactic sources; however, contrary to these expectations, at energies above 2 PeV, the \ic\ event rate drops, hinting at a steep decline in the incident neutrino flux itself at these multi-PeV energies. The theoretical challenge is to explain the apparent cutoff of the neutrino spectrum. 
The low number of observed events makes it difficult to conclusively determine the nature of the astrophysical sources responsible for the all-sky diffuse flux of neutrinos leading to these events. Several possible origins have been suggested, both astrophysical sources \cite{Anchordoqui:2013dnh, Anchordoqui:2013qsi, Laha:2013lka, Essey:2009ju, *Kalashev:2013vba, *Tjus:2014dna, Murase:2013rfa, Murase:2013ffa,Loeb:2006tw, *Tamborra:2014xia, *Anchordoqui:2014yva} and dark matter interactions \cite{Esmaili:2013gha, Bhattacharya:2014vwa, Feldstein:2013kka, Bai:2013nga}.

The role of slow-jet supernovae (SJS) as a possible source of UHE neutrino fluxes has been previously suggested in Refs.~\cite{Razzaque:2004yv, Razzaque:2005bh}, and has been explored in detail, see, \eg, \cite{Ando:2005xi, Horiuchi:2007xi}. SJS are core-collapse supernovae that have jets, similarly to gamma-ray bursts (GRB), although the jets in SJS have much lower Lorentz factors than the jets in a GRB and do not reach the envelope of the star. The environment is optically thick to photons and charged particles; therefore, the only visible sign of the jets may be the emitted neutrinos.

The neutrino flux produced from pion and kaon decays within these sources lies below the atmospheric neutrino background at TeV energies and beyond \cite{Razzaque:2004yv,Ando:2005xi}.
However, it has been shown \cite{Enberg:2008jm} that the decay of charmed \dmess\ (\dnot, \dnotb, \dpm) produced in $pp$ collisions within these sources leads to considerably higher neutrino fluxes with a spectrum resembling the shape of the proton flux, including the cut-off in the PeV energy range which is due to proton cooling processes starting to dominate over proton acceleration.
This previous estimate was done assuming an energy-independent, proton-spectrum weighted charm production cross-section \cite{Enberg:2008jm,Gandhi:2009qx}. 
In this paper, we show that the charm production even from each individual $pp$ collision in the source is significantly affected by the proton spectrum and its cut-off.
The effect of this on the proton-spectrum weighted moments of the charm production cross-section is crucial and modifies the previously considered \cite{Enberg:2008jm,Gandhi:2009qx} effect of the energy cut-off.
We demonstrate that, by incorporating this effect in the calculation of the diffuse flux from SJS sources, one could account for the observed IC excess events.

We consider the dependence of the diffuse flux on the uncertainties in the astrophysical parametrization of the source distribution in the universe as well as from theoretical uncertainties in the charm production cross-section from $pp$ collisions \cite{Vogt:2007aw}.

Apart from being sensitive to the proton spectral cutoff, the charm production cross section has theoretical uncertainties due to
\begin{inparaenum}[(\itshape a\upshape)]
	\item the different values of charm masses (we consider $m_c = 1.3\text{ and }1.5$ GeV)\footnote{As discussed in \cite{Vogt:2007aw}, the choice of charm mass $m_c = 1.5$ GeV seems to underestimate the total charm production cross-section seen at recent experiments, \eg, ALICE~\cite{ALICE:2011aa, *Abelev:2012vra}, LHCb~\cite{Aaij:2013mga} and ATLAS~\cite{ATLAS-CONF-2011-017}.}
	\item uncertainties in parton distribution functions and fragmentation functions, and
	\item different renormalization and factorization scales.
\end{inparaenum}
By considering the uncertainties in the relevant astrophysical parameters and in the charm cross-section, we compute the plausible range of variation of the resulting diffuse neutrino flux from SJS sources.
For reasonable choices of the theoretical parameters, we show that the diffuse neutrino flux coming from such sources could directly explain the most striking features of the flux that reproduces the IC events, namely
\begin{inparaenum}[\itshape 1\upshape)]
	\item the $E^{-2}$ behavior at energies 30 TeV--2 PeV consistent with the IC best-fit shown in Eq.\ \eqref{eqn:ic-bf}, and
	\item the drop in the flux at energies beyond 2 PeV.
\end{inparaenum}
The \ic\ events further seem to derive from an isotropic flux, with no clustering in neither time nor space. The diffuse flux naturally fulfills this requirement.

\section{\label{sec:indiv-source}Neutrino flux from charm decay in individual slow-jet source}
We consider neutrino production in the ``choked jets'' of mildly relativistic supernovae with bulk-Lorentz factor $\gamj \sim 3-10$ and a jet angle of $\theta_j \sim 1/\gamj$ \cite{Razzaque:2004yv}.
The number densities of electrons and protons in such sources are given by \cite{Razzaque:2005bh}
\begin{equation}
n^\prime_{e} = n^\prime_{p} = \frac{1}{2\pi m_{p}c^{3}}\frac{L_{j}}{(\theta_j r_{j} \gamj)^2}\,,
\label{eq:density}
\end{equation}
where primed quantities are given in the jet comoving frame. Here $L_j$ is the jet luminosity, $r_j = 2\gamj^2 c t_{v}$ is the jet radius, and $t_v$ is the jet variability time scale $\sim 0.1 \text{ s}$.
For our benchmark estimate, we use $L_j = 5\times 10^{50}\text{ erg s}^{-1}$ and $\gamj=4$.

Given the relatively high proton content in such slow jets, proton-proton collisions at high energies can dominate over $p\gamma$ interactions and lead to the production of $D$-mesons.
The decay lengths of $D$-mesons are much shorter than the corresponding interaction lengths, so they decay almost instantly, producing neutrinos (see, \eg, \cite{Enberg:2008jm}).

In the limit where meson decay dominates over meson cooling, the neutrino flux expected at Earth due to the decay of meson $M$, from a source at a distance $d_L$, is given by \cite{Enberg:2008jm}
\begin{equation}\label{eqn:phinu}
\phi_{\nu}(E) = Z_{M\nu} Z_{NM} \phi_{N}(E) \frac{L_j\gamj^2}{2\pi\theta_{j}^2 d_{L}^2 \log(E'_\text{max}/E'_\text{min})},
\end{equation}
where the spectrum-weighted moments $Z_{ij}$, discussed below, encode the production and decay of $D$-mesons. This expression does not hold when significant cooling of the meson occurs before it decays; for this case the full expressions from \cite{Enberg:2008jm} must be used.

The quantity  $\phi_{N}$ represents the proton flux within the source as seen from the Earth-observer frame, and $E'_\text{min} = m_p c^2$. The shape of the proton spectrum is dependent upon the shock acceleration parameters within the source.
As long as the protons take longer to cool (due to synchrotron radiation, inverse Compton scattering with thermal photons and interactions with hadrons or gammas) than to be accelerated to the particular energy, the proton spectrum is a power-law $\propto {E^{\prime}}^{-2}$.
Since acceleration times increase linearly with proton energies, as energies of the protons reach \order{1} PeV (in the comoving frame), the cooling times fall below the acceleration time, and with the cooling processes now dominating, the corresponding proton flux falls off steeply at higher energies.
The exact energy at which the crossover takes place depends on the specifics of the conditions inside the source.
The proton acceleration time is proportional to the energy and to $\kappa$, a parameter inversely related to the diffusion coefficient (see e.g.~\cite{Enberg:2008jm}).
We assume $\kappa=1$, implying
$E^\prime_\text{max} \approx 5 \times 10^{6}$ GeV in the jet-comoving frame, for our benchmark values of $L_j$ and \gamj.
Depending on the orientation of the magnetic fields, this coefficient could be as large as $\kappa=10$~\cite{Giacalone:2006yd,Protheroe:1998pj}, implying that the cut-off energy $E^\prime_\text{max}$ could be lower by about a factor of 10 from what we use here. We take the proton flux to be
$\phi_{N}(E^{\prime}) = A{E^\prime}^{-2}f_{N}(E^{\prime})$
with
\begin{equation}
f_{N}(E^{\prime}) = \left[ 1 + \left( \frac{E^{\prime}}{E^{\prime}_\text{max}} \right) \right] e^{-E^{\prime}/E^{\prime}_\text{max}}\,
\end{equation}
describing the cutoff behavior~\cite{Protheroe:1998pj}. 

In Eq.\ \eqref{eqn:phinu} $Z_{M\nu}$ and $Z_{NM}$ encode the decay of $M$ to neutrinos and the production of meson $M$ from $NN$ interactions, respectively. The $Z_{NM}$ are defined
as follows:
\begin{equation}
Z_{NM} = \int_0^1 \frac{\lambda_N(E)}{\lambda_N(E/x_E)}\frac{\phi_{N}(E/x_{E})}{\phi_{N}(E)}
\frac{\ud n_{N\to M}}{\ud x_{E}}
\frac{\diff x_{E}}{x_E}\,,
\end{equation}
where $\phi_{N}$ is the proton flux,
$x_E\equiv E_M/E_N$, $\lambda_N(E)$ is the hadronic cooling length for protons (see 
\eg~\cite{Enberg:2008jm}), and $dn/dx_E$ is the
energy distribution of the meson $M$ produced by $N$.
For our calculations, we take the $Z_{M\nu}$ values shown in Table II in \cite{Enberg:2008jm}.
We compute $Z_{pD}$ using the differential charm cross section calculated with the next-to-leading order $K$-factor following \cite{Pasquali:1998ji} but with CTEQ6.6 parton distributions~\cite{Nadolsky:2008zw}, and include fragmentation of charm quarks into charmed mesons.\footnote{In \cite{Enberg:2008jm} we used the charm cross section from the dipole picture calculation of \cite{Enberg:2008te}, but we have found \protect\cite{BERS} that this cross section has a slower growth with energy that falls below the recently measured charm cross section at the LHC~\cite{ALICE:2011aa, *Abelev:2012vra, Aaij:2013mga, ATLAS-CONF-2011-017}.}

We incorporate the energy dependence of the proton-flux in $Z_{pD}$ by including its energy cutoff factor $f_N$, and in addition $Z_{pD}$ depends on energy through the energy dependence of the charm total and differential cross sections.
We find that the mean $x_{E}$ for $pp \to D$ is around 0.2. Therefore, the most significant effect of incorporating the energy dependence in the $Z_{pD}$, is that the $D$-meson flux has a lower cutoff than the proton flux.

\begin{figure}[htb]
\begin{center}
	\includegraphics[width=\columnwidth]{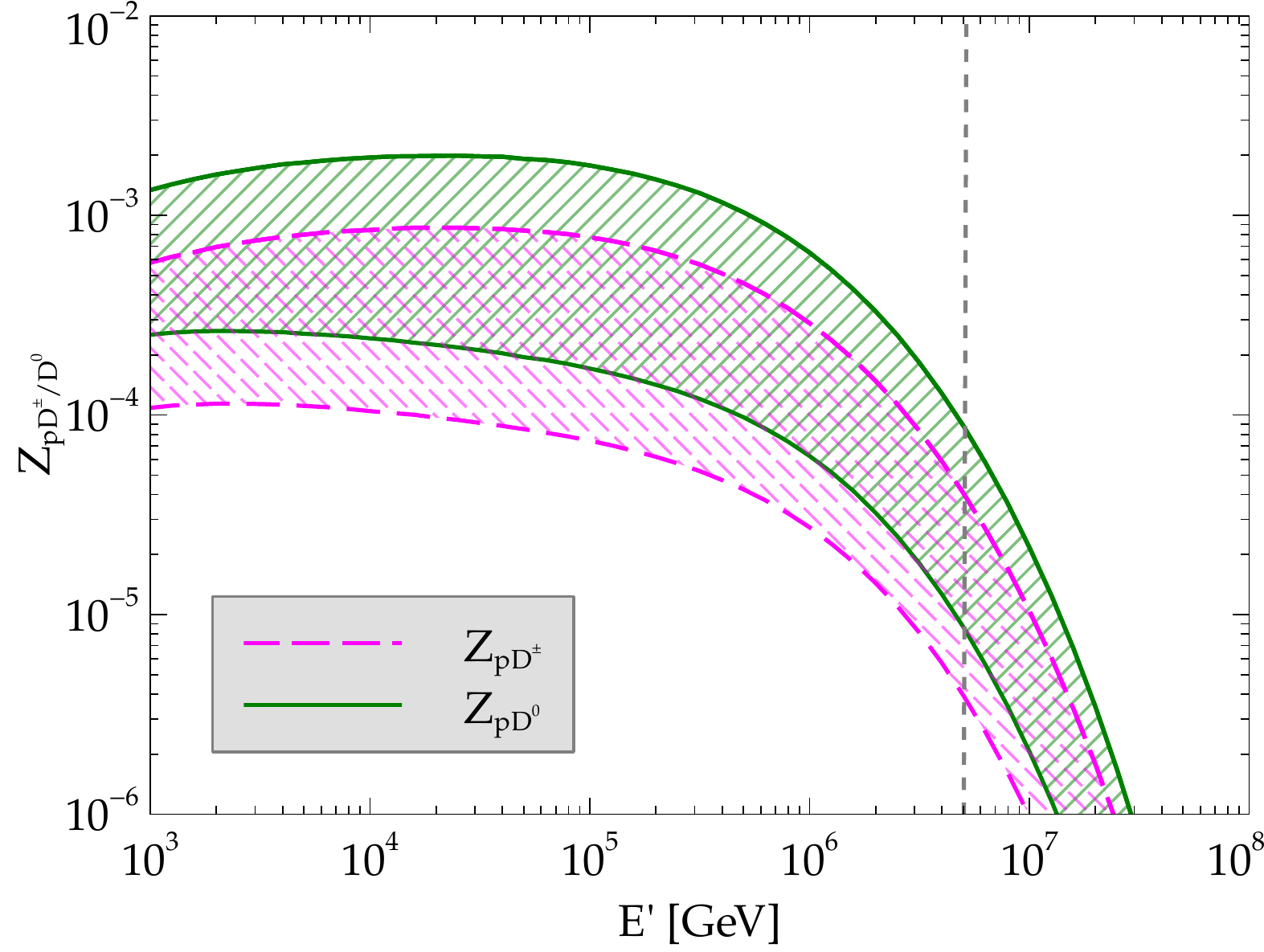}
\end{center}
\caption{
\label{fig:zmom}
$Z_{p\dnot}=Z_{p\dnotb}$ and $Z_{p\dpm}$ in the jet-comoving frame, shown with uncertainties due to variation in the assumed charm mass from $m_c = 1.3$ GeV to $m_c = 1.5$ GeV in each case, in addition to varying the renormalization ($\mu_R$) and factorization ($\mu_F$) scales within the range $[m_c, 2m_c]$.
	In each case, the lower limit corresponds to $m_c = 1.5$ GeV, $\mu_F = 2m_c$, $\mu_R = m_c$, while the upper limit corresponds to $m_c = 1.3$ GeV, $\mu_F = m_c$, $\mu_R = 2m_c$ respectively.
	The dotted vertical line indicates the $E_{\rm max}'$ for the proton flux in the source.
	}
\end{figure}

We present our results for the $Z$-moments in Fig.\ \ref{fig:zmom}. We show the uncertainty due to the choice of the charm mass, renormalization scale and factorization scale. The lower edge of the uncertainty band, with $m_c=1.5$ GeV, likely underestimates
the $Z$-moments. 

The $Z$-moments in Fig.\ \ref{fig:zmom} fall off at energies lower than the proton flux cut-off.
Previous estimates of the neutrino fluxes from the decay of $D$-mesons in slow-jet sources in the literature \cite{Enberg:2008jm,Gandhi:2009qx} were made assuming constant $Z_{pD}$, which, consequently, were larger in magnitude. This led to neutrino fluxes that closely mirrored the proton flux in spectral shape, which is not the case when accounting for the energy dependence in the $Z_{pD}$.

\section{\label{sec:diffuse-flux}Diffuse flux and IceCube}

The diffuse flux from SJS is given by (see, \eg, \cite{Razzaque:2004yv})
\begin{equation}\label{eqn:diff-flux}
\Phi(E_\nu) =
\frac{\xi_\text{sn}}{2\gamj^2}
\int_0^{\infty} \frac{\ndotsn(z)d_L^2 c t_j}{(1+z)^2}
\phi_\nu
\left[E_{\nu}(1+z)\right]
\bigabs{\frac{\ud t}{\ud z}}\diff z\,,
\end{equation}
where, $\ndotsn(z)$ represents the cosmic star-formation rate (\sfr) at a redshift $z$, $\phi_{\nu}$ represents the neutrino flux at energy $E_{\nu}$ from a single source, and the red-shift evolution of the universe is given by
\begin{equation}\label{eqn:dtdz}
\frac{\ud z}{\ud t} = H_{0} (1 + z) \sqrt{\OLam + \OM(1+z)^3}.
\end{equation}
The quantity
$t_j$ represents the burst duration of SJS, here assumed to be 10s, typical of such sources \cite{Razzaque:2004yv,Razzaque:2005bh}.
Following standard $\Lambda$CDM cosmology, the Hubble constant is $H_0 = 68 \hubbleunit$, and $\OM = 0.3175$, $\OLam = 0.6825$ \cite{Ade:2013zuv}, and
$0 \leqslant \xisn \leqslant 1$ represents the fraction of supernovae with slow-jets. Further, only a fraction $1/2\gamj^2$ of the total SJS population are directed toward Earth.
We use the \sfr\ modeled in \cite{Cucciati2011} for \ndotsn, with the normalization $\ndotsn(0) = 1.3 \times 10^{-4}\text{ yr}^{-1}\text{ Mpc}^{-3}$.

The number of events expected in IceCube from our predicted diffuse flux is computed by convoluting the diffuse flux with the effective neutrino area given in \cite{Aartsen:2013jdh}. The result is shown in Fig.~\ref{fig:diffeve}b, to be discussed below.

\section*{Astrophysical parameters and their uncertainties}

An earlier search by \ic\ for SJS puts limits on the bulk-Lorentz factor $\gamj$, the jet energy $E_{j}=L_{j}t_{j}$, and the fraction of SNe with jets \xisn~\cite{Abbasi:2011ja}.
For larger $\gamj\sim 10$, the limit on $\xi_\text{sn}$ is on the percent level for typical values of $E_{j}\sim 3-10 \times 10^{51}$~erg, but for $\gamj \lesssim 6$ these parameters are largely unconstrained.
Our benchmark values for these parameters are consistent with such observations.

Additionally, the local star formation rate has $1\sigma$ uncertainties itself, varying between $(1-2) \times 10^{-4}\text{ yr}^{-1}\text{ Mpc}^{-3}$ consistent with observed data \cite{Hopkins:2006bw, *Horiuchi:2011zz}.
These uncertainties in the astrophysical modeling of the sources add to the already existing uncertainties in the diffuse flux which originates from the charm cross-section.

While the various astrophysical parameters will ultimately be decided by observations, for our benchmark values, we see that the total diffuse neutrino-flux is consistent with IC observations for $\xisn\approx 1/3$.

\begin{figure*}
\begin{center}
\begin{tabular}{cc}
		\includegraphics[width=0.47\textwidth]{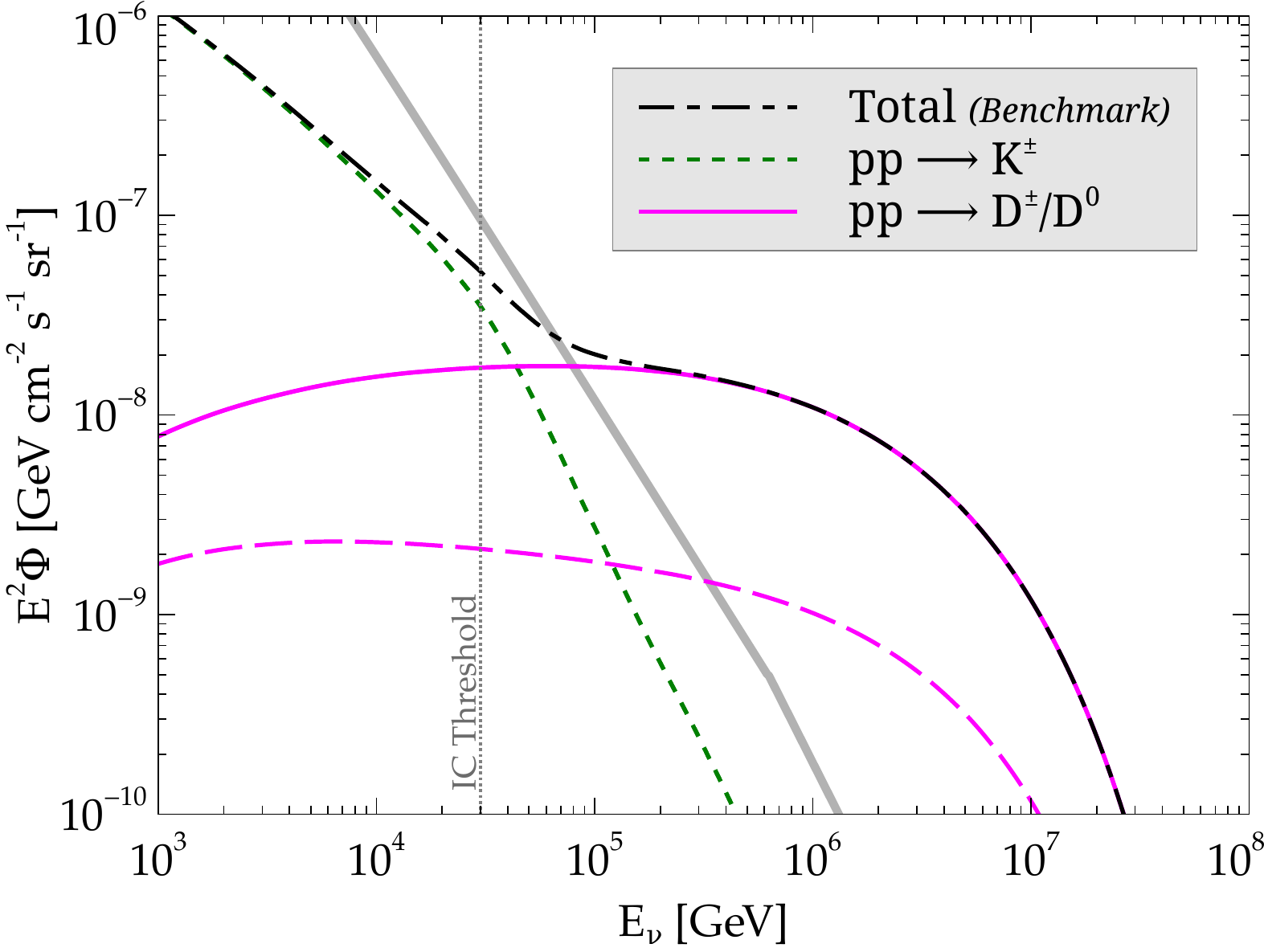} &
		\includegraphics[width=0.47\textwidth]{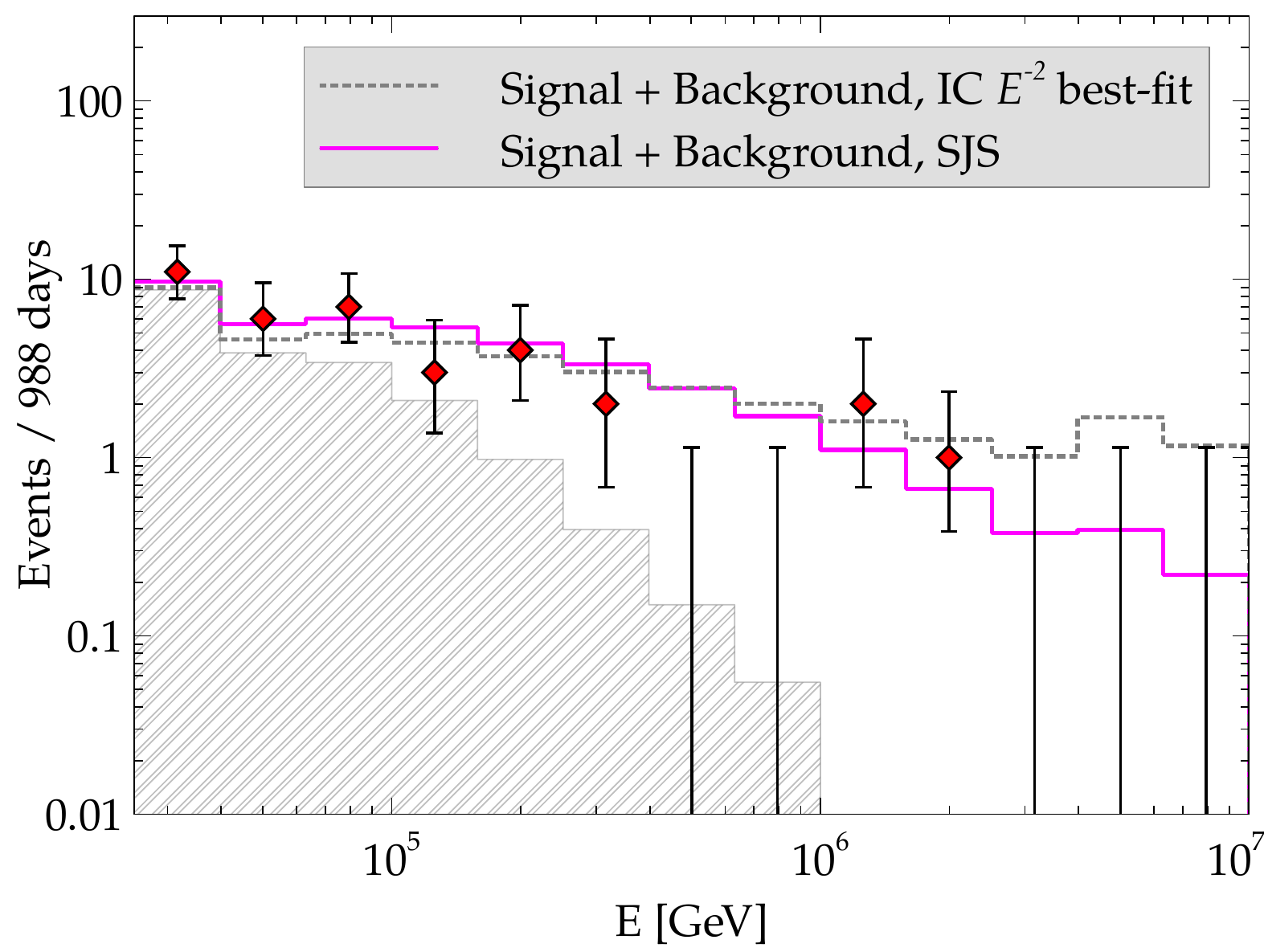} \\
		(a) & (b)
\end{tabular}
\end{center}	
	\caption{\label{fig:diffeve}
	(a) The benchmark neutrino ($\numu + \antinumu = \nue + \antinue$) diffuse flux (different production mechanisms as indicated in the legend) at earth (without accounting for neutrino oscillation), and (b) the corresponding estimated total event rates at IC (including neutrino flux from SJS after oscillation and atmospheric neutrino and muon background).
	The upper (solid) and lower (dashed) curves for the $D$-meson contributions correspond to different choices of the $m_c, \mu_R$ and $\mu_F$ discussed in Fig.\ \ref{fig:zmom}.
	In Fig.\ (a) the thick gray line represents the predicted (conventional) atmospheric $\numu + \bar{\nu}_\mu$ neutrino flux \cite{Gondolo:1995fq}.
	For the 988-day runtime for the IC, event-rate estimates [Fig.\ (b)] from the SJS benchmark  and IC $E^{-2}$ best-fit flux, and the observed IC event numbers \citep{Aartsen:2014gkd} (red diamonds) are shown.
	The mean event-rate estimates (light-gray shaded area) for background atmospheric neutrinos (including 90\% C.L.\ atmospheric charm contribution) and muons \cite{Aartsen:2014gkd} are also indicated.}
\end{figure*}

We show the diffuse neutrino flux produced from decay of kaons and charmed mesons in SJS sources in Fig.\ \ref{fig:diffeve}a.
The uncertainties in these fluxes due to differences in the assumed charm mass are also shown.
For our benchmark parameters, we find that the total flux from SJS is consistent with the IC best fit $E^{-2}$ flux up to about 2 PeV, while dropping off sharply thereafter, thus explaining the absence of events at the IC at multi-PeV energies.
In contrast, the flux from the decay of kaons is more steeply falling, and only contributes noticeably to the total flux at energies just above the IceCube threshold.
Using the total SJS diffuse flux (kaon + D-meson), we show the estimated 988-day event-rate at IC in Fig.\ \ref{fig:diffeve}b, comparing it to actual observations and against the event rates predicted from the IC best-fit $E^{-2}$ flux in Eq.\ \ref{eqn:ic-bf}.
\dmess\ decay to produce a predominantly $1:1:0$ ($\nue + \antinue : \numu + \antinumu : \nutau + \antinutau$) flavor composition of neutrinos at source, and we properly account for neutrino oscillation as they propagate to the earth while evaluating the predicted event rates at IC.
The corresponding neutrino mixing parameters are set at their present best-fit values \cite{Beringer:1900zz}.
The natural drop in the event rates beyond 2 PeV is consistent with the lack of events at $E \geqslant 2.1$ PeV at the IC, and in contrast to the $\sim 4$ events predicted by a uniform $E^{-2}$ flux with normalization set by Eq.\ \eqref{eqn:ic-bf}.

\section{\label{sec:conc}Conclusions}

We have shown that the recent UHE events seen at \ic\ are consistent with having their origin in a diffuse flux generated by decay of charmed mesons within the mildly relativistic jets of supernovae.
We have discussed the uncertainties in the magnitude and cutoff energies of the neutrino flux by considering the uncertainties in the D-meson production cross-section in $pp$ collisions and in the astrophysical parameters. We have included the energy dependence in the $Z$-moments for the computation of the $D$-meson production.
We find that for reasonable choices of the astrophysical and QCD parameters in the production of charm, the diffuse neutrino flux at Earth from such sources could be enough to explain the observed event rates at energies of 30 TeV to 2 PeV, while also exhibiting a sharp drop in the flux at energies above 2 PeV, in conformity with the lack of events at \ic\ at such high energies. QCD uncertainties in the charm production
cross section are large. Nevertheless,
for a range of parameters, the neutrino flux from D-meson decays within slow jet astrophysical sources could form a significant component of the total diffuse flux seen at \ic.

With more \ic\ events, it should be possible to
ascertain if the observed neutrinos indeed originate from charmed meson decays because of the
distinctive cutoff-like spectral nature of the flux.
If, in the future, \ic\ were to find that the diffuse flux were consistent with an unbroken power-law spectrum even at energies beyond 2 PeV and extending into the tens of PeV's, the slow-jet supernovae charmed-meson-origin hypothesis of the incoming neutrino flux would be disfavored.

\begin{acknowledgments}
This research was supported in part by the US Department of Energy contracts DE-FG02-04ER41319, DE-FG02-04ER41298, DE-FG03-91ER40662, DE-FG02-13ER41976 and DE-SC0010114, and in part by the Swedish Research Council under contract 621-2011-5107.

\end{acknowledgments}

\bibliography{sjs_ic_3}

\begin{thebibliography}{43}%
\makeatletter
\providecommand \@ifxundefined [1]{%
 \@ifx{#1\undefined}
}%
\providecommand \@ifnum [1]{%
 \ifnum #1\expandafter \@firstoftwo
 \else \expandafter \@secondoftwo
 \fi
}%
\providecommand \@ifx [1]{%
 \ifx #1\expandafter \@firstoftwo
 \else \expandafter \@secondoftwo
 \fi
}%
\providecommand \natexlab [1]{#1}%
\providecommand \enquote  [1]{``#1''}%
\providecommand \bibnamefont  [1]{#1}%
\providecommand \bibfnamefont [1]{#1}%
\providecommand \citenamefont [1]{#1}%
\providecommand \href@noop [0]{\@secondoftwo}%
\providecommand \href [0]{\begingroup \@sanitize@url \@href}%
\providecommand \@href[1]{\@@startlink{#1}\@@href}%
\providecommand \@@href[1]{\endgroup#1\@@endlink}%
\providecommand \@sanitize@url [0]{\catcode `\\12\catcode `\$12\catcode
  `\&12\catcode `\#12\catcode `\^12\catcode `\_12\catcode `\%12\relax}%
\providecommand \@@startlink[1]{}%
\providecommand \@@endlink[0]{}%
\providecommand \url  [0]{\begingroup\@sanitize@url \@url }%
\providecommand \@url [1]{\endgroup\@href {#1}{\urlprefix }}%
\providecommand \urlprefix  [0]{URL }%
\providecommand \Eprint [0]{\href }%
\providecommand \doibase [0]{http://dx.doi.org/}%
\providecommand \selectlanguage [0]{\@gobble}%
\providecommand \bibinfo  [0]{\@secondoftwo}%
\providecommand \bibfield  [0]{\@secondoftwo}%
\providecommand \translation [1]{[#1]}%
\providecommand \BibitemOpen [0]{}%
\providecommand \bibitemStop [0]{}%
\providecommand \bibitemNoStop [0]{.\EOS\space}%
\providecommand \EOS [0]{\spacefactor3000\relax}%
\providecommand \BibitemShut  [1]{\csname bibitem#1\endcsname}%
\let\auto@bib@innerbib\@empty
\bibitem [{\citenamefont {Aartsen}\ \emph
  {et~al.}(2013{\natexlab{a}})\citenamefont {Aartsen} \emph
  {et~al.}}]{Aartsen:2013bka}%
  \BibitemOpen
  \bibfield  {author} {\bibinfo {author} {\bibfnamefont {M.}~\bibnamefont
  {Aartsen}} \emph {et~al.} (\bibinfo {collaboration} {IceCube
  Collaboration}),\ }\href {\doibase 10.1103/PhysRevLett.111.021103} {\bibfield
   {journal} {\bibinfo  {journal} {Phys.Rev.Lett.}\ }\textbf {\bibinfo {volume}
  {111}},\ \bibinfo {pages} {021103} (\bibinfo {year} {2013}{\natexlab{a}})},\
  \Eprint {http://arxiv.org/abs/1304.5356} {arXiv:1304.5356 [astro-ph.HE]}
  \BibitemShut {NoStop}%
\bibitem [{\citenamefont {Aartsen}\ \emph
  {et~al.}(2013{\natexlab{b}})\citenamefont {Aartsen} \emph
  {et~al.}}]{Aartsen:2013jdh}%
  \BibitemOpen
  \bibfield  {author} {\bibinfo {author} {\bibfnamefont {M.}~\bibnamefont
  {Aartsen}} \emph {et~al.} (\bibinfo {collaboration} {IceCube}),\ }\href
  {\doibase 10.1126/science.1242856} {\bibfield  {journal} {\bibinfo  {journal}
  {Science}\ }\textbf {\bibinfo {volume} {342}},\ \bibinfo {pages} {1242856}
  (\bibinfo {year} {2013}{\natexlab{b}})},\ \Eprint
  {http://arxiv.org/abs/1311.5238} {arXiv:1311.5238 [astro-ph.HE]} \BibitemShut
  {NoStop}%
\bibitem [{\citenamefont {Aartsen}\ \emph {et~al.}(2014)\citenamefont {Aartsen}
  \emph {et~al.}}]{Aartsen:2014gkd}%
  \BibitemOpen
  \bibfield  {author} {\bibinfo {author} {\bibfnamefont {M.}~\bibnamefont
  {Aartsen}} \emph {et~al.} (\bibinfo {collaboration} {IceCube
  Collaboration}),\ }\href@noop {} {\  (\bibinfo {year} {2014})},\ \Eprint
  {http://arxiv.org/abs/1405.5303} {arXiv:1405.5303 [astro-ph.HE]} \BibitemShut
  {NoStop}%
\bibitem [{\citenamefont {Van~Elewyck}(2014)}]{VanElewyck:2013xja}%
  \BibitemOpen
  \bibfield  {author} {\bibinfo {author} {\bibfnamefont {V.}~\bibnamefont
  {Van~Elewyck}} (\bibinfo {collaboration} {ANTARES}),\ }\href {\doibase
  10.1016/j.nima.2013.11.092} {\bibfield  {journal} {\bibinfo  {journal}
  {Nucl.Instrum.Meth.}\ }\textbf {\bibinfo {volume} {A742}},\ \bibinfo {pages}
  {63} (\bibinfo {year} {2014})},\ \Eprint {http://arxiv.org/abs/1311.7002}
  {arXiv:1311.7002 [astro-ph.HE]} \BibitemShut {NoStop}%
\bibitem [{\citenamefont {Anchordoqui}\ \emph
  {et~al.}(2014{\natexlab{a}})\citenamefont {Anchordoqui}, \citenamefont
  {Barger}, \citenamefont {Cholis}, \citenamefont {Goldberg}, \citenamefont
  {Hooper} \emph {et~al.}}]{Anchordoqui:2013dnh}%
  \BibitemOpen
  \bibfield  {author} {\bibinfo {author} {\bibfnamefont {L.~A.}\ \bibnamefont
  {Anchordoqui}}, \bibinfo {author} {\bibfnamefont {V.}~\bibnamefont {Barger}},
  \bibinfo {author} {\bibfnamefont {I.}~\bibnamefont {Cholis}}, \bibinfo
  {author} {\bibfnamefont {H.}~\bibnamefont {Goldberg}}, \bibinfo {author}
  {\bibfnamefont {D.}~\bibnamefont {Hooper}},  \emph {et~al.},\ }\href
  {\doibase 10.1016/j.jheap.2014.01.001} {\bibfield  {journal} {\bibinfo
  {journal} {Journal of High Energy Astrophysics}\ }\textbf {\bibinfo {volume}
  {1-2}},\ \bibinfo {pages} {1} (\bibinfo {year} {2014}{\natexlab{a}})},\
  \Eprint {http://arxiv.org/abs/1312.6587} {arXiv:1312.6587 [astro-ph.HE]}
  \BibitemShut {NoStop}%
\bibitem [{\citenamefont {Anchordoqui}\ \emph
  {et~al.}(2014{\natexlab{b}})\citenamefont {Anchordoqui}, \citenamefont
  {Goldberg}, \citenamefont {Lynch}, \citenamefont {Olinto}, \citenamefont
  {Paul} \emph {et~al.}}]{Anchordoqui:2013qsi}%
  \BibitemOpen
  \bibfield  {author} {\bibinfo {author} {\bibfnamefont {L.~A.}\ \bibnamefont
  {Anchordoqui}}, \bibinfo {author} {\bibfnamefont {H.}~\bibnamefont
  {Goldberg}}, \bibinfo {author} {\bibfnamefont {M.~H.}\ \bibnamefont {Lynch}},
  \bibinfo {author} {\bibfnamefont {A.~V.}\ \bibnamefont {Olinto}}, \bibinfo
  {author} {\bibfnamefont {T.~C.}\ \bibnamefont {Paul}},  \emph {et~al.},\
  }\href@noop {} {\bibfield  {journal} {\bibinfo  {journal} {Phys.Rev.}\
  }\textbf {\bibinfo {volume} {D89}},\ \bibinfo {pages} {083003} (\bibinfo
  {year} {2014}{\natexlab{b}})},\ \Eprint {http://arxiv.org/abs/1306.5021}
  {arXiv:1306.5021 [astro-ph.HE]} \BibitemShut {NoStop}%
\bibitem [{\citenamefont {Laha}\ \emph {et~al.}(2013)\citenamefont {Laha},
  \citenamefont {Beacom}, \citenamefont {Dasgupta}, \citenamefont {Horiuchi},\
  and\ \citenamefont {Murase}}]{Laha:2013lka}%
  \BibitemOpen
  \bibfield  {author} {\bibinfo {author} {\bibfnamefont {R.}~\bibnamefont
  {Laha}}, \bibinfo {author} {\bibfnamefont {J.~F.}\ \bibnamefont {Beacom}},
  \bibinfo {author} {\bibfnamefont {B.}~\bibnamefont {Dasgupta}}, \bibinfo
  {author} {\bibfnamefont {S.}~\bibnamefont {Horiuchi}}, \ and\ \bibinfo
  {author} {\bibfnamefont {K.}~\bibnamefont {Murase}},\ }\href {\doibase
  10.1103/PhysRevD.88.043009} {\bibfield  {journal} {\bibinfo  {journal}
  {Phys.Rev.}\ }\textbf {\bibinfo {volume} {D88}},\ \bibinfo {pages} {043009}
  (\bibinfo {year} {2013})},\ \Eprint {http://arxiv.org/abs/1306.2309}
  {arXiv:1306.2309 [astro-ph.HE]} \BibitemShut {NoStop}%
\bibitem [{\citenamefont {Essey}\ \emph {et~al.}(2010)\citenamefont {Essey},
  \citenamefont {Kalashev}, \citenamefont {Kusenko},\ and\ \citenamefont
  {Beacom}}]{Essey:2009ju}%
  \BibitemOpen
  \bibfield  {author} {\bibinfo {author} {\bibfnamefont {W.}~\bibnamefont
  {Essey}}, \bibinfo {author} {\bibfnamefont {O.~E.}\ \bibnamefont {Kalashev}},
  \bibinfo {author} {\bibfnamefont {A.}~\bibnamefont {Kusenko}}, \ and\
  \bibinfo {author} {\bibfnamefont {J.~F.}\ \bibnamefont {Beacom}},\ }\href
  {\doibase 10.1103/PhysRevLett.104.141102} {\bibfield  {journal} {\bibinfo
  {journal} {Phys.Rev.Lett.}\ }\textbf {\bibinfo {volume} {104}},\ \bibinfo
  {pages} {141102} (\bibinfo {year} {2010})},\ \Eprint
  {http://arxiv.org/abs/0912.3976} {arXiv:0912.3976 [astro-ph.HE]} \BibitemShut
  {NoStop}%
\bibitem [{\citenamefont {Kalashev}\ \emph {et~al.}(2013)\citenamefont
  {Kalashev}, \citenamefont {Kusenko},\ and\ \citenamefont
  {Essey}}]{Kalashev:2013vba}%
  \BibitemOpen
  \bibfield  {author} {\bibinfo {author} {\bibfnamefont {O.~E.}\ \bibnamefont
  {Kalashev}}, \bibinfo {author} {\bibfnamefont {A.}~\bibnamefont {Kusenko}}, \
  and\ \bibinfo {author} {\bibfnamefont {W.}~\bibnamefont {Essey}},\ }\href
  {\doibase 10.1103/PhysRevLett.111.041103} {\bibfield  {journal} {\bibinfo
  {journal} {Phys.Rev.Lett.}\ }\textbf {\bibinfo {volume} {111}},\ \bibinfo
  {pages} {041103} (\bibinfo {year} {2013})},\ \Eprint
  {http://arxiv.org/abs/1303.0300} {arXiv:1303.0300 [astro-ph.HE]} \BibitemShut
  {NoStop}%
\bibitem [{\citenamefont {Tjus}\ \emph {et~al.}(2014)\citenamefont {Tjus},
  \citenamefont {Eichmann}, \citenamefont {Halzen}, \citenamefont
  {Kheirandish},\ and\ \citenamefont {Saba}}]{Tjus:2014dna}%
  \BibitemOpen
  \bibfield  {author} {\bibinfo {author} {\bibfnamefont {J.~B.}\ \bibnamefont
  {Tjus}}, \bibinfo {author} {\bibfnamefont {B.}~\bibnamefont {Eichmann}},
  \bibinfo {author} {\bibfnamefont {F.}~\bibnamefont {Halzen}}, \bibinfo
  {author} {\bibfnamefont {A.}~\bibnamefont {Kheirandish}}, \ and\ \bibinfo
  {author} {\bibfnamefont {S.}~\bibnamefont {Saba}},\ }\href@noop {} {\
  (\bibinfo {year} {2014})},\ \Eprint {http://arxiv.org/abs/1406.0506}
  {arXiv:1406.0506 [astro-ph.HE]} \BibitemShut {NoStop}%
\bibitem [{\citenamefont {Murase}\ \emph {et~al.}(2013)\citenamefont {Murase},
  \citenamefont {Ahlers},\ and\ \citenamefont {Lacki}}]{Murase:2013rfa}%
  \BibitemOpen
  \bibfield  {author} {\bibinfo {author} {\bibfnamefont {K.}~\bibnamefont
  {Murase}}, \bibinfo {author} {\bibfnamefont {M.}~\bibnamefont {Ahlers}}, \
  and\ \bibinfo {author} {\bibfnamefont {B.~C.}\ \bibnamefont {Lacki}},\ }\href
  {\doibase 10.1103/PhysRevD.88.121301} {\bibfield  {journal} {\bibinfo
  {journal} {Phys.Rev.}\ }\textbf {\bibinfo {volume} {D88}},\ \bibinfo {pages}
  {121301} (\bibinfo {year} {2013})},\ \Eprint {http://arxiv.org/abs/1306.3417}
  {arXiv:1306.3417 [astro-ph.HE]} \BibitemShut {NoStop}%
\bibitem [{\citenamefont {Murase}\ and\ \citenamefont
  {Ioka}(2013)}]{Murase:2013ffa}%
  \BibitemOpen
  \bibfield  {author} {\bibinfo {author} {\bibfnamefont {K.}~\bibnamefont
  {Murase}}\ and\ \bibinfo {author} {\bibfnamefont {K.}~\bibnamefont {Ioka}},\
  }\href {\doibase 10.1103/PhysRevLett.111.121102} {\bibfield  {journal}
  {\bibinfo  {journal} {Phys.Rev.Lett.}\ }\textbf {\bibinfo {volume} {111}},\
  \bibinfo {pages} {121102} (\bibinfo {year} {2013})},\ \Eprint
  {http://arxiv.org/abs/1306.2274} {arXiv:1306.2274 [astro-ph.HE]} \BibitemShut
  {NoStop}%
\bibitem [{\citenamefont {Loeb}\ and\ \citenamefont
  {Waxman}(2006)}]{Loeb:2006tw}%
  \BibitemOpen
  \bibfield  {author} {\bibinfo {author} {\bibfnamefont {A.}~\bibnamefont
  {Loeb}}\ and\ \bibinfo {author} {\bibfnamefont {E.}~\bibnamefont {Waxman}},\
  }\href {\doibase 10.1088/1475-7516/2006/05/003} {\bibfield  {journal}
  {\bibinfo  {journal} {JCAP}\ }\textbf {\bibinfo {volume} {0605}},\ \bibinfo
  {pages} {003} (\bibinfo {year} {2006})},\ \Eprint
  {http://arxiv.org/abs/astro-ph/0601695} {arXiv:astro-ph/0601695 [astro-ph]}
  \BibitemShut {NoStop}%
\bibitem [{\citenamefont {Tamborra}\ \emph {et~al.}(2014)\citenamefont
  {Tamborra}, \citenamefont {Ando},\ and\ \citenamefont
  {Murase}}]{Tamborra:2014xia}%
  \BibitemOpen
  \bibfield  {author} {\bibinfo {author} {\bibfnamefont {I.}~\bibnamefont
  {Tamborra}}, \bibinfo {author} {\bibfnamefont {S.}~\bibnamefont {Ando}}, \
  and\ \bibinfo {author} {\bibfnamefont {K.}~\bibnamefont {Murase}},\
  }\href@noop {} {\  (\bibinfo {year} {2014})},\ \Eprint
  {http://arxiv.org/abs/1404.1189} {arXiv:1404.1189 [astro-ph.HE]} \BibitemShut
  {NoStop}%
\bibitem [{\citenamefont {Anchordoqui}\ \emph
  {et~al.}(2014{\natexlab{c}})\citenamefont {Anchordoqui}, \citenamefont
  {Paul}, \citenamefont {da~Silva}, \citenamefont {Torres},\ and\ \citenamefont
  {Vlcek}}]{Anchordoqui:2014yva}%
  \BibitemOpen
  \bibfield  {author} {\bibinfo {author} {\bibfnamefont {L.~A.}\ \bibnamefont
  {Anchordoqui}}, \bibinfo {author} {\bibfnamefont {T.~C.}\ \bibnamefont
  {Paul}}, \bibinfo {author} {\bibfnamefont {L.~H.~M.}\ \bibnamefont
  {da~Silva}}, \bibinfo {author} {\bibfnamefont {D.~F.}\ \bibnamefont
  {Torres}}, \ and\ \bibinfo {author} {\bibfnamefont {B.~J.}\ \bibnamefont
  {Vlcek}},\ }\href@noop {} {\  (\bibinfo {year} {2014}{\natexlab{c}})},\
  \Eprint {http://arxiv.org/abs/1405.7648} {arXiv:1405.7648 [astro-ph.HE]}
  \BibitemShut {NoStop}%
\bibitem [{\citenamefont {Esmaili}\ and\ \citenamefont
  {Serpico}(2013)}]{Esmaili:2013gha}%
  \BibitemOpen
  \bibfield  {author} {\bibinfo {author} {\bibfnamefont {A.}~\bibnamefont
  {Esmaili}}\ and\ \bibinfo {author} {\bibfnamefont {P.~D.}\ \bibnamefont
  {Serpico}},\ }\href {\doibase 10.1088/1475-7516/2013/11/054} {\bibfield
  {journal} {\bibinfo  {journal} {JCAP}\ }\textbf {\bibinfo {volume} {1311}},\
  \bibinfo {pages} {054} (\bibinfo {year} {2013})},\ \Eprint
  {http://arxiv.org/abs/1308.1105} {arXiv:1308.1105 [hep-ph]} \BibitemShut
  {NoStop}%
\bibitem [{\citenamefont {Bhattacharya}\ \emph
  {et~al.}(2014{\natexlab{a}})\citenamefont {Bhattacharya}, \citenamefont
  {Reno},\ and\ \citenamefont {Sarcevic}}]{Bhattacharya:2014vwa}%
  \BibitemOpen
  \bibfield  {author} {\bibinfo {author} {\bibfnamefont {A.}~\bibnamefont
  {Bhattacharya}}, \bibinfo {author} {\bibfnamefont {M.~H.}\ \bibnamefont
  {Reno}}, \ and\ \bibinfo {author} {\bibfnamefont {I.}~\bibnamefont
  {Sarcevic}},\ }\href {\doibase 10.1007/JHEP06(2014)110} {\bibfield  {journal}
  {\bibinfo  {journal} {JHEP}\ }\textbf {\bibinfo {volume} {06}},\ \bibinfo
  {pages} {110} (\bibinfo {year} {2014}{\natexlab{a}})},\ \Eprint
  {http://arxiv.org/abs/1403.1862} {arXiv:1403.1862 [hep-ph]} \BibitemShut
  {NoStop}%
\bibitem [{\citenamefont {Feldstein}\ \emph {et~al.}(2013)\citenamefont
  {Feldstein}, \citenamefont {Kusenko}, \citenamefont {Matsumoto},\ and\
  \citenamefont {Yanagida}}]{Feldstein:2013kka}%
  \BibitemOpen
  \bibfield  {author} {\bibinfo {author} {\bibfnamefont {B.}~\bibnamefont
  {Feldstein}}, \bibinfo {author} {\bibfnamefont {A.}~\bibnamefont {Kusenko}},
  \bibinfo {author} {\bibfnamefont {S.}~\bibnamefont {Matsumoto}}, \ and\
  \bibinfo {author} {\bibfnamefont {T.~T.}\ \bibnamefont {Yanagida}},\ }\href
  {\doibase 10.1103/PhysRevD.88.015004} {\bibfield  {journal} {\bibinfo
  {journal} {Phys.Rev.}\ }\textbf {\bibinfo {volume} {D88}},\ \bibinfo {pages}
  {015004} (\bibinfo {year} {2013})},\ \Eprint {http://arxiv.org/abs/1303.7320}
  {arXiv:1303.7320 [hep-ph]} \BibitemShut {NoStop}%
\bibitem [{\citenamefont {Bai}\ \emph {et~al.}(2013)\citenamefont {Bai},
  \citenamefont {Lu},\ and\ \citenamefont {Salvado}}]{Bai:2013nga}%
  \BibitemOpen
  \bibfield  {author} {\bibinfo {author} {\bibfnamefont {Y.}~\bibnamefont
  {Bai}}, \bibinfo {author} {\bibfnamefont {R.}~\bibnamefont {Lu}}, \ and\
  \bibinfo {author} {\bibfnamefont {J.}~\bibnamefont {Salvado}},\ }\href@noop
  {} {\  (\bibinfo {year} {2013})},\ \Eprint {http://arxiv.org/abs/1311.5864}
  {arXiv:1311.5864 [hep-ph]} \BibitemShut {NoStop}%
\bibitem [{\citenamefont {Razzaque}\ \emph {et~al.}(2004)\citenamefont
  {Razzaque}, \citenamefont {Meszaros},\ and\ \citenamefont
  {Waxman}}]{Razzaque:2004yv}%
  \BibitemOpen
  \bibfield  {author} {\bibinfo {author} {\bibfnamefont {S.}~\bibnamefont
  {Razzaque}}, \bibinfo {author} {\bibfnamefont {P.}~\bibnamefont {Meszaros}},
  \ and\ \bibinfo {author} {\bibfnamefont {E.}~\bibnamefont {Waxman}},\ }\href
  {\doibase 10.1103/PhysRevLett.94.109903} {\bibfield  {journal} {\bibinfo
  {journal} {Phys.Rev.Lett.}\ }\textbf {\bibinfo {volume} {93}},\ \bibinfo
  {pages} {181101} (\bibinfo {year} {2004})},\ \Eprint
  {http://arxiv.org/abs/astro-ph/0407064} {arXiv:astro-ph/0407064 [astro-ph]}
  \BibitemShut {NoStop}%
\bibitem [{\citenamefont {Razzaque}\ \emph {et~al.}(2005)\citenamefont
  {Razzaque}, \citenamefont {Meszaros},\ and\ \citenamefont
  {Waxman}}]{Razzaque:2005bh}%
  \BibitemOpen
  \bibfield  {author} {\bibinfo {author} {\bibfnamefont {S.}~\bibnamefont
  {Razzaque}}, \bibinfo {author} {\bibfnamefont {P.}~\bibnamefont {Meszaros}},
  \ and\ \bibinfo {author} {\bibfnamefont {E.}~\bibnamefont {Waxman}},\ }\href
  {\doibase 10.1142/S0217732305018414} {\bibfield  {journal} {\bibinfo
  {journal} {Mod.Phys.Lett.}\ }\textbf {\bibinfo {volume} {A20}},\ \bibinfo
  {pages} {2351} (\bibinfo {year} {2005})},\ \Eprint
  {http://arxiv.org/abs/astro-ph/0509729} {arXiv:astro-ph/0509729 [astro-ph]}
  \BibitemShut {NoStop}%
\bibitem [{\citenamefont {Ando}\ and\ \citenamefont
  {Beacom}(2005)}]{Ando:2005xi}%
  \BibitemOpen
  \bibfield  {author} {\bibinfo {author} {\bibfnamefont {S.}~\bibnamefont
  {Ando}}\ and\ \bibinfo {author} {\bibfnamefont {J.~F.}\ \bibnamefont
  {Beacom}},\ }\href {\doibase 10.1103/PhysRevLett.95.061103} {\bibfield
  {journal} {\bibinfo  {journal} {Phys.Rev.Lett.}\ }\textbf {\bibinfo {volume}
  {95}},\ \bibinfo {pages} {061103} (\bibinfo {year} {2005})},\ \Eprint
  {http://arxiv.org/abs/astro-ph/0502521} {arXiv:astro-ph/0502521 [astro-ph]}
  \BibitemShut {NoStop}%
\bibitem [{\citenamefont {Horiuchi}\ and\ \citenamefont
  {Ando}(2008)}]{Horiuchi:2007xi}%
  \BibitemOpen
  \bibfield  {author} {\bibinfo {author} {\bibfnamefont {S.}~\bibnamefont
  {Horiuchi}}\ and\ \bibinfo {author} {\bibfnamefont {S.}~\bibnamefont
  {Ando}},\ }\href {\doibase 10.1103/PhysRevD.77.063007} {\bibfield  {journal}
  {\bibinfo  {journal} {Phys.Rev.}\ }\textbf {\bibinfo {volume} {D77}},\
  \bibinfo {pages} {063007} (\bibinfo {year} {2008})},\ \Eprint
  {http://arxiv.org/abs/0711.2580} {arXiv:0711.2580 [astro-ph]} \BibitemShut
  {NoStop}%
\bibitem [{\citenamefont {Enberg}\ \emph {et~al.}(2009)\citenamefont {Enberg},
  \citenamefont {Reno},\ and\ \citenamefont {Sarcevic}}]{Enberg:2008jm}%
  \BibitemOpen
  \bibfield  {author} {\bibinfo {author} {\bibfnamefont {R.}~\bibnamefont
  {Enberg}}, \bibinfo {author} {\bibfnamefont {M.~H.}\ \bibnamefont {Reno}}, \
  and\ \bibinfo {author} {\bibfnamefont {I.}~\bibnamefont {Sarcevic}},\ }\href
  {\doibase 10.1103/PhysRevD.79.053006} {\bibfield  {journal} {\bibinfo
  {journal} {Phys.Rev.}\ }\textbf {\bibinfo {volume} {D79}},\ \bibinfo {pages}
  {053006} (\bibinfo {year} {2009})},\ \Eprint {http://arxiv.org/abs/0808.2807}
  {arXiv:0808.2807 [astro-ph]} \BibitemShut {NoStop}%
\bibitem [{\citenamefont {Gandhi}\ \emph {et~al.}(2009)\citenamefont {Gandhi},
  \citenamefont {Samanta},\ and\ \citenamefont {Watanabe}}]{Gandhi:2009qx}%
  \BibitemOpen
  \bibfield  {author} {\bibinfo {author} {\bibfnamefont {R.}~\bibnamefont
  {Gandhi}}, \bibinfo {author} {\bibfnamefont {A.}~\bibnamefont {Samanta}}, \
  and\ \bibinfo {author} {\bibfnamefont {A.}~\bibnamefont {Watanabe}},\ }\href
  {\doibase 10.1088/1475-7516/2009/09/015} {\bibfield  {journal} {\bibinfo
  {journal} {JCAP}\ }\textbf {\bibinfo {volume} {0909}},\ \bibinfo {pages}
  {015} (\bibinfo {year} {2009})},\ \Eprint {http://arxiv.org/abs/0905.2483}
  {arXiv:0905.2483 [hep-ph]} \BibitemShut {NoStop}%
\bibitem [{\citenamefont {Vogt}(2008)}]{Vogt:2007aw}%
  \BibitemOpen
  \bibfield  {author} {\bibinfo {author} {\bibfnamefont {R.}~\bibnamefont
  {Vogt}},\ }\href {\doibase 10.1140/epjst/e2008-00603-5} {\bibfield  {journal}
  {\bibinfo  {journal} {Eur.Phys.J.ST}\ }\textbf {\bibinfo {volume} {155}},\
  \bibinfo {pages} {213} (\bibinfo {year} {2008})},\ \Eprint
  {http://arxiv.org/abs/0709.2531} {arXiv:0709.2531 [hep-ph]} \BibitemShut
  {NoStop}%
\bibitem [{\citenamefont {Abelev}\ \emph
  {et~al.}(2012{\natexlab{a}})\citenamefont {Abelev} \emph
  {et~al.}}]{ALICE:2011aa}%
  \BibitemOpen
  \bibfield  {author} {\bibinfo {author} {\bibfnamefont {B.}~\bibnamefont
  {Abelev}} \emph {et~al.} (\bibinfo {collaboration} {ALICE Collaboration}),\
  }\href {\doibase 10.1007/JHEP01(2012)128} {\bibfield  {journal} {\bibinfo
  {journal} {JHEP}\ }\textbf {\bibinfo {volume} {1201}},\ \bibinfo {pages}
  {128} (\bibinfo {year} {2012}{\natexlab{a}})},\ \Eprint
  {http://arxiv.org/abs/1111.1553} {arXiv:1111.1553 [hep-ex]} \BibitemShut
  {NoStop}%
\bibitem [{\citenamefont {Abelev}\ \emph
  {et~al.}(2012{\natexlab{b}})\citenamefont {Abelev} \emph
  {et~al.}}]{Abelev:2012vra}%
  \BibitemOpen
  \bibfield  {author} {\bibinfo {author} {\bibfnamefont {B.}~\bibnamefont
  {Abelev}} \emph {et~al.} (\bibinfo {collaboration} {ALICE Collaboration}),\
  }\href {\doibase 10.1007/JHEP07(2012)191} {\bibfield  {journal} {\bibinfo
  {journal} {JHEP}\ }\textbf {\bibinfo {volume} {1207}},\ \bibinfo {pages}
  {191} (\bibinfo {year} {2012}{\natexlab{b}})},\ \Eprint
  {http://arxiv.org/abs/1205.4007} {arXiv:1205.4007 [hep-ex]} \BibitemShut
  {NoStop}%
\bibitem [{\citenamefont {Aaij}\ \emph {et~al.}(2013)\citenamefont {Aaij} \emph
  {et~al.}}]{Aaij:2013mga}%
  \BibitemOpen
  \bibfield  {author} {\bibinfo {author} {\bibfnamefont {R.}~\bibnamefont
  {Aaij}} \emph {et~al.} (\bibinfo {collaboration} {LHCb collaboration}),\
  }\href {\doibase 10.1016/j.nuclphysb.2013.02.010} {\bibfield  {journal}
  {\bibinfo  {journal} {Nucl.Phys.}\ }\textbf {\bibinfo {volume} {B871}},\
  \bibinfo {pages} {1} (\bibinfo {year} {2013})},\ \Eprint
  {http://arxiv.org/abs/1302.2864} {arXiv:1302.2864 [hep-ex]} \BibitemShut
  {NoStop}%
\bibitem [{\citenamefont {{ATLAS Collaboration}}(2011)}]{ATLAS-CONF-2011-017}%
  \BibitemOpen
  \bibfield  {author} {\bibinfo {author} {\bibnamefont {{ATLAS
  Collaboration}}},\ }\href@noop {} {\emph {\bibinfo {title} {{Measurement of
  $D^{(*)}$ meson production cross sections in pp collisions at $\sqrt{s}=7$
  TeV with the ATLAS detector}}}},\ \bibinfo {type} {Tech. Rep.}\ \bibinfo
  {number} {ATLAS-CONF-2011-017}\ (\bibinfo  {institution} {CERN},\ \bibinfo
  {address} {Geneva},\ \bibinfo {year} {2011})\BibitemShut {NoStop}%
\bibitem [{\citenamefont {Giacalone}\ and\ \citenamefont
  {Jokipii}(2006)}]{Giacalone:2006yd}%
  \BibitemOpen
  \bibfield  {author} {\bibinfo {author} {\bibfnamefont {J.}~\bibnamefont
  {Giacalone}}\ and\ \bibinfo {author} {\bibfnamefont {J.}~\bibnamefont
  {Jokipii}},\ }\href {\doibase 10.1088/1742-6596/47/1/020} {\bibfield
  {journal} {\bibinfo  {journal} {J.Phys.Conf.Ser.}\ }\textbf {\bibinfo
  {volume} {47}},\ \bibinfo {pages} {160} (\bibinfo {year} {2006})}\BibitemShut
  {NoStop}%
\bibitem [{\citenamefont {Protheroe}\ and\ \citenamefont
  {Stanev}(1999)}]{Protheroe:1998pj}%
  \BibitemOpen
  \bibfield  {author} {\bibinfo {author} {\bibfnamefont {R.}~\bibnamefont
  {Protheroe}}\ and\ \bibinfo {author} {\bibfnamefont {T.}~\bibnamefont
  {Stanev}},\ }\href {\doibase 10.1016/S0927-6505(98)00055-3} {\bibfield
  {journal} {\bibinfo  {journal} {Astropart.Phys.}\ }\textbf {\bibinfo {volume}
  {10}},\ \bibinfo {pages} {185} (\bibinfo {year} {1999})},\ \Eprint
  {http://arxiv.org/abs/astro-ph/9808129} {arXiv:astro-ph/9808129 [astro-ph]}
  \BibitemShut {NoStop}%
\bibitem [{\citenamefont {Pasquali}\ \emph {et~al.}(1999)\citenamefont
  {Pasquali}, \citenamefont {Reno},\ and\ \citenamefont
  {Sarcevic}}]{Pasquali:1998ji}%
  \BibitemOpen
  \bibfield  {author} {\bibinfo {author} {\bibfnamefont {L.}~\bibnamefont
  {Pasquali}}, \bibinfo {author} {\bibfnamefont {M.}~\bibnamefont {Reno}}, \
  and\ \bibinfo {author} {\bibfnamefont {I.}~\bibnamefont {Sarcevic}},\ }\href
  {\doibase 10.1103/PhysRevD.59.034020} {\bibfield  {journal} {\bibinfo
  {journal} {Phys.Rev.}\ }\textbf {\bibinfo {volume} {D59}},\ \bibinfo {pages}
  {034020} (\bibinfo {year} {1999})},\ \Eprint
  {http://arxiv.org/abs/hep-ph/9806428} {arXiv:hep-ph/9806428 [hep-ph]}
  \BibitemShut {NoStop}%
\bibitem [{\citenamefont {Nadolsky}\ \emph {et~al.}(2008)\citenamefont
  {Nadolsky}, \citenamefont {Lai}, \citenamefont {Cao}, \citenamefont {Huston},
  \citenamefont {Pumplin} \emph {et~al.}}]{Nadolsky:2008zw}%
  \BibitemOpen
  \bibfield  {author} {\bibinfo {author} {\bibfnamefont {P.~M.}\ \bibnamefont
  {Nadolsky}}, \bibinfo {author} {\bibfnamefont {H.-L.}\ \bibnamefont {Lai}},
  \bibinfo {author} {\bibfnamefont {Q.-H.}\ \bibnamefont {Cao}}, \bibinfo
  {author} {\bibfnamefont {J.}~\bibnamefont {Huston}}, \bibinfo {author}
  {\bibfnamefont {J.}~\bibnamefont {Pumplin}},  \emph {et~al.},\ }\href
  {\doibase 10.1103/PhysRevD.78.013004} {\bibfield  {journal} {\bibinfo
  {journal} {Phys.Rev.}\ }\textbf {\bibinfo {volume} {D78}},\ \bibinfo {pages}
  {013004} (\bibinfo {year} {2008})},\ \Eprint {http://arxiv.org/abs/0802.0007}
  {arXiv:0802.0007 [hep-ph]} \BibitemShut {NoStop}%
\bibitem [{\citenamefont {Enberg}\ \emph {et~al.}(2008)\citenamefont {Enberg},
  \citenamefont {Reno},\ and\ \citenamefont {Sarcevic}}]{Enberg:2008te}%
  \BibitemOpen
  \bibfield  {author} {\bibinfo {author} {\bibfnamefont {R.}~\bibnamefont
  {Enberg}}, \bibinfo {author} {\bibfnamefont {M.~H.}\ \bibnamefont {Reno}}, \
  and\ \bibinfo {author} {\bibfnamefont {I.}~\bibnamefont {Sarcevic}},\ }\href
  {\doibase 10.1103/PhysRevD.78.043005} {\bibfield  {journal} {\bibinfo
  {journal} {Phys.Rev.}\ }\textbf {\bibinfo {volume} {D78}},\ \bibinfo {pages}
  {043005} (\bibinfo {year} {2008})},\ \Eprint {http://arxiv.org/abs/0806.0418}
  {arXiv:0806.0418 [hep-ph]} \BibitemShut {NoStop}%
\bibitem [{\citenamefont {Bhattacharya}\ \emph
  {et~al.}(2014{\natexlab{b}})\citenamefont {Bhattacharya}, \citenamefont
  {Enberg}, \citenamefont {Reno},\ and\ \citenamefont {Sarcevic}}]{BERS}%
  \BibitemOpen
  \bibfield  {author} {\bibinfo {author} {\bibfnamefont {A.}~\bibnamefont
  {Bhattacharya}}, \bibinfo {author} {\bibfnamefont {R.}~\bibnamefont
  {Enberg}}, \bibinfo {author} {\bibfnamefont {M.~H.}\ \bibnamefont {Reno}}, \
  and\ \bibinfo {author} {\bibfnamefont {I.}~\bibnamefont {Sarcevic}},\
  }\href@noop {} {} (\bibinfo {year} {2014}{\natexlab{b}}),\ \bibinfo {note}
  {{in preparation}}\BibitemShut {NoStop}%
\bibitem [{\citenamefont {Ade}\ \emph {et~al.}(2013)\citenamefont {Ade} \emph
  {et~al.}}]{Ade:2013zuv}%
  \BibitemOpen
  \bibfield  {author} {\bibinfo {author} {\bibfnamefont {P.}~\bibnamefont
  {Ade}} \emph {et~al.} (\bibinfo {collaboration} {Planck Collaboration}),\
  }\href@noop {} {\  (\bibinfo {year} {2013})},\ \Eprint
  {http://arxiv.org/abs/1303.5076} {arXiv:1303.5076 [astro-ph.CO]} \BibitemShut
  {NoStop}%
\bibitem [{\citenamefont {{Cucciati, O.}}\ \emph {et~al.}(2012)\citenamefont
  {{Cucciati, O.}} \emph {et~al.}}]{Cucciati2011}%
  \BibitemOpen
  \bibfield  {author} {\bibinfo {author} {\bibnamefont {{Cucciati, O.}}} \emph
  {et~al.},\ }\href {\doibase 10.1051/0004-6361/201118010} {\bibfield
  {journal} {\bibinfo  {journal} {A\&A}\ }\textbf {\bibinfo {volume} {539}},\
  \bibinfo {pages} {A31} (\bibinfo {year} {2012})},\ \Eprint
  {http://arxiv.org/abs/1109.1005} {arXiv:1109.1005 [astro-ph.CO]} \BibitemShut
  {NoStop}%
\bibitem [{\citenamefont {Abbasi}\ \emph {et~al.}(2012)\citenamefont {Abbasi}
  \emph {et~al.}}]{Abbasi:2011ja}%
  \BibitemOpen
  \bibfield  {author} {\bibinfo {author} {\bibfnamefont {R.}~\bibnamefont
  {Abbasi}} \emph {et~al.} (\bibinfo {collaboration} {IceCube Collaboration,
  ROTSE Collaboration}),\ }\href {\doibase 10.1051/0004-6361/201118071}
  {\bibfield  {journal} {\bibinfo  {journal} {Astron.Astrophys.}\ }\textbf
  {\bibinfo {volume} {539}},\ \bibinfo {pages} {A60} (\bibinfo {year}
  {2012})},\ \Eprint {http://arxiv.org/abs/1111.7030} {arXiv:1111.7030
  [astro-ph.HE]} \BibitemShut {NoStop}%
\bibitem [{\citenamefont {Hopkins}\ and\ \citenamefont
  {Beacom}(2006)}]{Hopkins:2006bw}%
  \BibitemOpen
  \bibfield  {author} {\bibinfo {author} {\bibfnamefont {A.~M.}\ \bibnamefont
  {Hopkins}}\ and\ \bibinfo {author} {\bibfnamefont {J.~F.}\ \bibnamefont
  {Beacom}},\ }\href {\doibase 10.1086/506610} {\bibfield  {journal} {\bibinfo
  {journal} {Astrophys.J.}\ }\textbf {\bibinfo {volume} {651}},\ \bibinfo
  {pages} {142} (\bibinfo {year} {2006})},\ \Eprint
  {http://arxiv.org/abs/astro-ph/0601463} {arXiv:astro-ph/0601463 [astro-ph]}
  \BibitemShut {NoStop}%
\bibitem [{\citenamefont {Horiuchi}\ \emph {et~al.}(2011)\citenamefont
  {Horiuchi}, \citenamefont {Beacom}, \citenamefont {Kochanek}, \citenamefont
  {Prieto}, \citenamefont {Stanek} \emph {et~al.}}]{Horiuchi:2011zz}%
  \BibitemOpen
  \bibfield  {author} {\bibinfo {author} {\bibfnamefont {S.}~\bibnamefont
  {Horiuchi}}, \bibinfo {author} {\bibfnamefont {J.~F.}\ \bibnamefont
  {Beacom}}, \bibinfo {author} {\bibfnamefont {C.~S.}\ \bibnamefont
  {Kochanek}}, \bibinfo {author} {\bibfnamefont {J.~L.}\ \bibnamefont
  {Prieto}}, \bibinfo {author} {\bibfnamefont {K.}~\bibnamefont {Stanek}},
  \emph {et~al.},\ }\href {\doibase 10.1088/0004-637X/738/2/154} {\bibfield
  {journal} {\bibinfo  {journal} {Astrophys.J.}\ }\textbf {\bibinfo {volume}
  {738}},\ \bibinfo {pages} {154} (\bibinfo {year} {2011})},\ \Eprint
  {http://arxiv.org/abs/1102.1977} {arXiv:1102.1977 [astro-ph.CO]} \BibitemShut
  {NoStop}%
\bibitem [{\citenamefont {Gondolo}\ \emph {et~al.}(1996)\citenamefont
  {Gondolo}, \citenamefont {Ingelman},\ and\ \citenamefont
  {Thunman}}]{Gondolo:1995fq}%
  \BibitemOpen
  \bibfield  {author} {\bibinfo {author} {\bibfnamefont {P.}~\bibnamefont
  {Gondolo}}, \bibinfo {author} {\bibfnamefont {G.}~\bibnamefont {Ingelman}}, \
  and\ \bibinfo {author} {\bibfnamefont {M.}~\bibnamefont {Thunman}},\ }\href
  {\doibase 10.1016/0927-6505(96)00033-3} {\bibfield  {journal} {\bibinfo
  {journal} {Astropart.Phys.}\ }\textbf {\bibinfo {volume} {5}},\ \bibinfo
  {pages} {309} (\bibinfo {year} {1996})},\ \Eprint
  {http://arxiv.org/abs/hep-ph/9505417} {arXiv:hep-ph/9505417 [hep-ph]}
  \BibitemShut {NoStop}%
\bibitem [{\citenamefont {Beringer}\ \emph {et~al.}(2012)\citenamefont
  {Beringer} \emph {et~al.}}]{Beringer:1900zz}%
  \BibitemOpen
  \bibfield  {author} {\bibinfo {author} {\bibfnamefont {J.}~\bibnamefont
  {Beringer}} \emph {et~al.} (\bibinfo {collaboration} {Particle Data Group}),\
  }\href {\doibase 10.1103/PhysRevD.86.010001} {\bibfield  {journal} {\bibinfo
  {journal} {Phys.Rev.}\ }\textbf {\bibinfo {volume} {D86}},\ \bibinfo {pages}
  {010001} (\bibinfo {year} {2012})}\BibitemShut {NoStop}%
\end{thebibliography}%
\end{document}